\documentclass[lettersize,journal]{IEEEtran}

\usepackage[compact]{titlesec}
\usepackage{newtxtext,newtxmath}
\usepackage{booktabs}
\usepackage{adjustbox}
\usepackage{subcaption}
\usepackage{multirow}
\usepackage{xcolor}
\usepackage{breqn,xspace}
\usepackage{bm}
\usepackage{lipsum,multicol}
\usepackage{makecell}
\usepackage{booktabs}
\usepackage{algorithm}
\usepackage{algorithmic}
\usepackage{amsmath}
\usepackage{url}

\ifodd 0
\newcommand{\rev}[1]{{\color{blue}#1}}
\newcommand{\newrev}[1]{{\color{red}#1}}

\else
\newcommand{\rev}[1]{#1}
\newcommand{\newrev}[1]{#1}

\fi

\newcommand{\name}{DVA\xspace}

\begin{document}

\title{LEO Satellite Networks Assisted Geo-distributed Data Processing}

\author{Zhiyuan Zhao, Zhe Chen,~\IEEEmembership{Member,~IEEE}, Zheng Lin, Wenjun Zhu, Kun Qiu, Chaoqun You and Yue Gao,~\IEEEmembership{Fellow,~IEEE}
\thanks{Z. Zhao, Z. Chen, Z. Lin, W. Zhu, and Y. Gao are with the School of Computer Science, Fudan University, Shanghai 200438, China (e-mail: zhaozhiyuan256@gmail.com; zhechen@fudan.edu.cn; zlin20@fudan.edu.cn; wenjun@fudan.edu.cn; chaoqunyou@fudan.edu.cn; qkun@fudan.edu.cn; gao.yue@fudan.edu.cn). (Corresponding author: Yue Gao)}
}
\maketitle

\begin{abstract}

Nowadays, the increasing deployment of edge clouds globally provides users with low-latency services. However, connecting an edge cloud to a core cloud via optic cables in terrestrial networks poses significant barriers due to the prohibitively expensive building cost of optic cables. Fortunately, emerging Low Earth Orbit~(LEO) satellite networks~(e.g., Starlink) offer a more cost-effective solution for increasing edge clouds, and hence large volumes of data in edge clouds can be transferred to a core cloud via those networks for time-sensitive big data tasks processing, such as attack detection. \newrev{However, the state-of-the-art satellite selection algorithms bring poor performance for those processing via our measurements.} Therefore, we propose a novel data volume aware satellite selection algorithm, named \name, to support such big data processing tasks. \rev{\name first takes into account both the data size in edge clouds and satellite capacity to finalize the selection, thereby preventing congestion in the access network and reducing \newrev{transmitting duration}.} Extensive simulations validate that \name has a significantly lower average access network duration than the state-of-the-art satellite selection algorithms in a LEO satellite emulation platform.

% Extensive experiments demonstrate that \name outperforms the state-of-the-art satellite selection algorithms in an LEO satellite emulation platform.

% The increasing deployment of edge clouds globally provides users with low-latency services, but it is non-trivial to connect an edge cloud to a core cloud via optic cables in terrestrial networks, since the building cost of optic cables is prohibitively expensive. Fortunately, emerging Low Earth Orbit~(LEO) satellite networks~(e.g., Starlink) provide a more cost-effective solution for increasing edge clouds, and hence large data in edge clouds can be transferred to a core cloud by them for time-sensitive task processing~(e.g., attack detection). However, we find that the state-of-the-art satellite selection algorithms bring bad performance for those processing via our measurements. We propose a novel satellite selection algorithm, called \name to support such data processing tasks. \rev{\name first takes into account both the data size in edge clouds and satellite capacity to finalize the selection, preventing congestion in the access network and reducing transmission duration.} Finally, we demonstrate \name outperforms the state-of-the-art satellite selection algorithms in an LEO satellite emulation platform.
\end{abstract}


\section{Introduction}

\newrev{
Internet service providers such as Google and Amazon 
}
are extensively deploying edge clouds worldwide to provide users with low-latency service request responses~\cite{shen2022edgematrix,liu2024collaborative}. 
% \newrev{Internet} companies like Google and Amazon have global telecommunications networks and construct thousands of edge clouds worldwide to provide real-time computing services to local users. 
For instance, Google Distributed Cloud Edge provides solutions for the retail industry, including sales terminals and self-checkout systems, and Amazon utilizes CloudFront to offer users high-speed data caching and content delivery services. Establishing edge clouds requires network connectivity to data centers, and the cost of deploying fiber optic networks between edge clouds and data centers is relatively high. On average, the installation cost of fiber optics ranges from \textdollar10,000 to \textdollar30,000 per kilometer~\cite{genuinemodulesMuchDoes}. If the distance between edge and cloud is considerable or the terrain is challenging, the deployment cost may be even higher. This makes widespread deployment of edge clouds across different regions challenging.

\begin{figure}[!t]
\centering
\includegraphics[width=3.2 in]{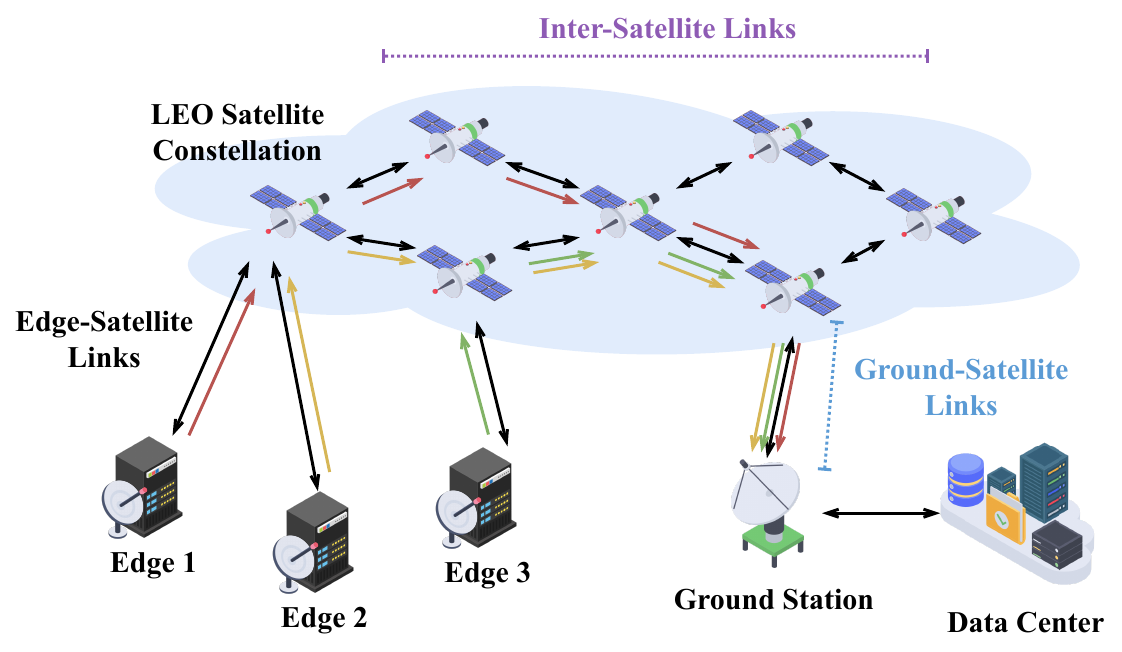}
\caption{Edge clouds access the data center via satellite network.}
\label{figure:1}
\end{figure}

In recent years, the emergence of Low Earth Orbit~(LEO) satellite networks has greatly expanded the coverage of broadband internet~\cite{li2020user,al2021session,wu2023split}. Companies like Starlink have proposed their own LEO satellite constellation plans, with tens of thousands of satellites set to be deployed in LEO~\cite{yuan2023graph,lin2023fedsn,yuan2024satsense}. This initiative aims to provide high-bandwidth, low-latency network transmission to every corner of the Earth~\cite{li2020user}. LEO also offers a more cost-effective option for building new edge clouds and connecting them to the cloud, as illustrated in Fig.~\ref{figure:1} \newrev{leveraging} the broadband access services provided by LEO satellites eliminates the need for expensive ground-based fiber optic infrastructure. This enables edge clouds to communicate with data centers from anywhere, thus extending computational capabilities to more locations in need. To expand the coverage of AWS services, Amazon has initiated the Kuiper project, which includes thousands of satellites, and extensively constructing ground stations adjacent to data centers~\cite{kodheli2020satellite}. \rev{Kuiper plans to provide cloud computing service in the areas where terrestrial networks cannot reach.}
% what do you mean?

While satellite networks facilitate easier access to edge clouds, this deployment method also presents many underexplored issues. One of the scenarios of interest in this paper is the processing of big data. Edge clouds deployed in various regions continuously generate a vast amount of data. Some of this data requires periodic or real-time transmission from edge clouds to data centers for processing~\cite{wang2022vabus,su2023primal}. This processing is necessary to promptly complete tasks such as analysis, prediction, and dynamic maintenance. For example, network operation logs from various edge nodes require periodic transmission to data centers for potential attack detection. Scientific observation data needs real-time aggregation and processing for scientific analysis. Additionally, log data serving users in certain regions requires real-time processing for dynamic content delivery. The real-time nature of these requirements necessitates that data from edge nodes be transmitted to data centers as quickly as possible, thereby reducing the time from data generation to output results.

\begin{figure}[!t]
\centering
\includegraphics[width=2.8 in]{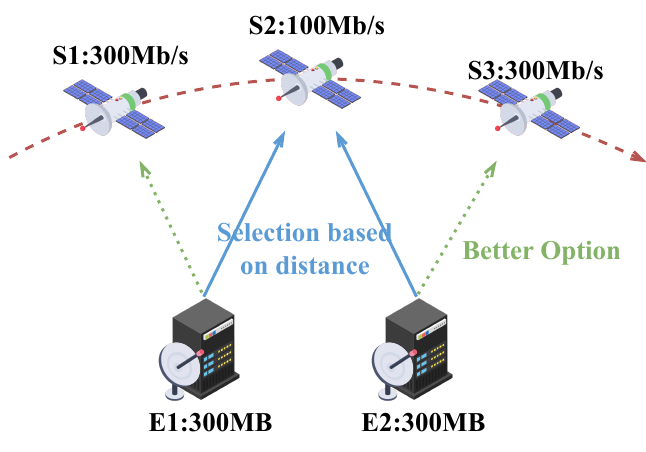}
\caption{The default selection algorithm may lead to unnecessary bandwidth competition.}
\label{figure:2}
\end{figure}

In the scenario of using LEO satellites for edge cloud data communication, the access network's bandwidth constitutes the bottleneck for transmission. Taking Starlink as an example, the order of magnitude for the user access bandwidth, ground station access bandwidth, and inter-satellite link bandwidth are approximately \rev{0.5~\!Gb/s}, \rev{2.5~\!Gb/s}, and \rev{20 Gb/s}, respectively~\cite{del2019technical}. Therefore, selecting the appropriate access satellite for each edge node participating in data processing tasks is crucial for reducing transmission duration. The default access satellite selection strategy typically relies solely on the relative positions of edge nodes and satellites as the matching criterion~\cite{liu2022enabling}. For instance, in the shortest distance-based satellite selection algorithm, the satellite closest to the sender is chosen as the access node, similarly, in the longest visible time-based satellite selection algorithm, the satellite expected to remain within the sender's field of view for the longest duration is selected as the access node. As shown in Fig.~\ref{figure:2}, these selection schemes do not take into account the data transmission volume at the network level or the current available bandwidth capacity of satellites. This oversight can lead to unnecessary competition for network bandwidth resources, hindering the realization of maximum throughput and minimizing transmission duration potential.

Our work introduces a satellite access matching algorithm called \name, designed specifically for distributed edge cloud big data real-time centralized processing scenarios. \name aims to match each edge cloud participating in data processing tasks with the most suitable access satellite. \name leverages the predictable position characteristics of LEO satellite constellations to compute a set of available access satellites for each edge cloud. It then matches edge nodes to satellites based on data volume and available satellite bandwidth, aiming to minimize transmission duration and maximize throughput for the access network portion. Based on the greedy algorithm's principles, \name has a relatively low computational time complexity, ensuring timely completion of matching schemes and prompt initiation of the transmission process.

Based on simulation experiments conducted using satellite simulation tools like STK, we have demonstrated that \name, compared to the baseline default selection algorithm, reduces the average access network transmission duration by approximately 50\% and increases access network throughput by more than double on average. Compared to the theoretical optimal algorithm obtained through modeling with integer linear programming and solved by the Gurobi optimizer, \name reduces the computation duration from approximately 290 ms to around 1 ms, while \newrev{guarantee} that the transmission duration does not exceed 1.1 times the optimal result.

The contributions of this letter are as follows:
\begin{itemize}
\item{Discovery and utilization of the mechanism by which edge clouds access data centers via satellite networks.}
\item{Design of corresponding satellite selection algorithms, reducing the transmission duration of data from edge clouds to data centers and maximizing the available bandwidth of the access network.}
\item{Design and simulation experiments to validate the effectiveness of the algorithms.}
\end{itemize}

The structure of this letter is as follows: Sec.~\ref{sec:design} introduces the design of the algorithm, including problem modeling, design principles, and specific steps. Section~\ref{sec:evaluation} presents the simulation experiments designed to validate the effectiveness of the algorithm, comparison with other algorithms, and the experimental results. Section~\ref{sec:conclusion} summarizes the article.

\section{Algorithm Design} \label{sec:design}
In this section, we will provide a detailed description of the overall design and algorithmic process of \name. We will introduce the design principles of the algorithm and analyze its time complexity.

As shown in Fig.~\ref{figure:1}, the \name system consists of three components: data center, edge cloud nodes deployed in various locations, and a LEO satellite network.

\begin{itemize}
\item{The data center serves as the control core of the system, maintaining information such as the positions and data volumes of edge clouds. Additionally, it can obtain real-time movement status of the LEO satellite constellation from satellite operators.}
\item{The edge clouds are distributed across various regions globally and run a variety of services, such as user-facing content services and scientific data collection services. For the data sets of interest, edge clouds periodically report their data status to the data center, including information about data set types, sizes, and other relevant details.}
\item{The LEO satellite network comprises a constellation of satellites covering the entire globe, providing network access services to edge clouds. Through periodic measurements conducted by satellite network operators, real-time data on the available satellite bandwidth can be obtained.}
\end{itemize}

The process of selecting access satellites generally includes three steps. Firstly, calculating a candidate satellite set based on the relative positions of the edge cloud and satellites. Secondly, computing matching results based on data volume and available bandwidth.
Lastly, publishing the matching results and initiating data transmission. Below, we will provide a detailed explanation of the problem modeling and algorithmic process.

\subsection{Calculation of Candidate Satellite Constellations}

\begin{figure}[!t]
\centering
\includegraphics[width=2.8 in]{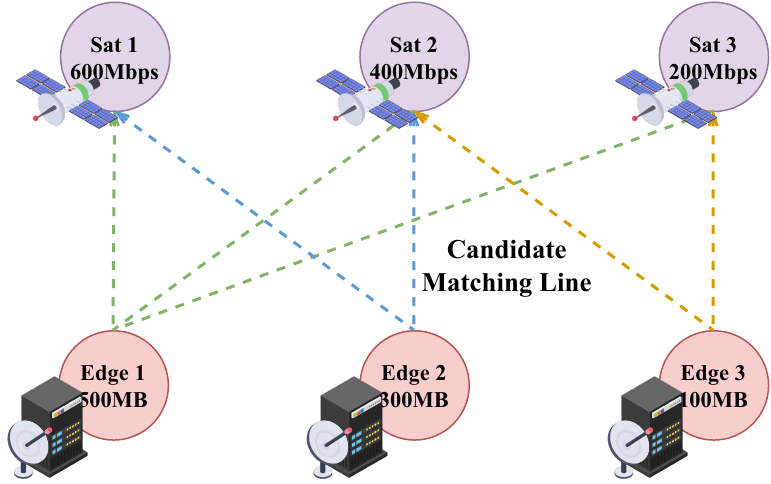}
\caption{Edge - candidate access satellite bipartite graph.}
\label{figure:3}
\end{figure}

During the candidate satellite constellation calculation phase, the data center computes the available satellites within the field of view of each edge node based on their geographical coordinates and Satellite Two-Line Elements (TLE) files. This process results in a bipartite graph containing both the edge nodes and the satellites, as illustrated in Fig.~\ref{figure:3}. On the bottom side of the graph, nodes represent edge clouds participating in data processing tasks, including information about the data volume associated with each edge cloud. On the top side, nodes represent available access satellites, including current available bandwidth data for each satellite. The lines connecting the two sets of nodes represent potential matches, indicating that the satellite on the top can serve as an access point for the edge cloud on the bottom.

From the diagram, it is evident that there exists a many-to-many relationship between edge clouds and satellites. An edge cloud can select multiple satellites within its field of view, and likewise, a satellite may be chosen by multiple edge clouds simultaneously. With the default selection algorithm, there is a possibility of mistakenly choosing satellites with lower carrying capacity due to the neglect of satellite bandwidth. Similarly, multiple edge clouds may simultaneously select the same satellite, thereby exacerbating unnecessary competition for network resources.

\subsection{Problem Formulation}
After obtaining the bipartite graph representing the edge clouds and candidate access satellites, the next step is mathematical modeling. Since only the access network is considered, the satellite selection algorithm does not include the data center in the model.

Let \( E = \{e_1, e_2, \ldots, e_m\} \) denote the set of edge cloud nodes, comprising a total of \( m \) nodes participating in data processing tasks. \( D = \{d_1, d_2, \ldots, d_m\} \) represents the data volume that needs to be transmitted by each edge cloud. Let \( S = \{s_1, s_2, \ldots, s_n\} \) denote the set of candidate access satellites, comprising \( n \) candidate satellites in total. \( C = \{c_1, c_2, \ldots, c_n\} \) represents the current available bandwidth of each candidate access satellite. Let \( v_{a,b} \) represent the visibility relationship between edge cloud \( a \) and candidate satellite \( b \). If \( v_{a,b} = 1 \), it indicates that \( b \) can serve as an access satellite for \( a \). In the bipartite graph, a line connecting \( a \) and \( b \) can be identified. Let \( x_{a,b} \) denote the decision variable for satellite selection, which can only take on the values 0 or 1. Setting \( x_{a,b} \) to 1 indicates selecting satellite \( b \) as the access satellite for edge cloud \( a \).

Let \( T \) denote the data transmission time for the access network. Then, \( T \) is equal to the time required for the access satellite that takes the longest to receive all data from the edge clouds. Then, for any edge cloud node index \( i \) (where \( 1 \leq i \leq m \)) and any chosen satellite index \( j \) (where \( 1 \leq i \leq n \)), the following constraint holds:

\begin{equation}
\frac{\sum_{i} d_i \cdot x_{i,j}}{c_j} \leq T
\end{equation}

The objective of the model is to minimize the transmission duration \( T \) by assigning appropriate values to the decision variables \( x \). This model can be represented as an integer linear programming (ILP) model as follows:

\begin{gather}
\text{min} \quad T \\
\text{Subject to:} \quad \forall i \in [1, m], \forall j \in [1, n], \quad x_{i,j} \leq v_{i,j} \\
\forall i \in [1, m], \quad \sum_{j=1}^{n} x_{i,j} = 1
\end{gather}

Due to the NP-hard nature of the integer linear programming problem, obtaining the optimal solution requires enumerating every possible matching scenario. When dealing with small-scale edge node or satellite constellation data, results can be obtained relatively quickly. However, as the data scale increases, the time complexity of enumeration can reach \(O(m^n)\), rendering it unable to meet the real-time requirements of satellite selection. Therefore, the design of heuristic algorithms becomes necessary. These algorithms aim to compute better matching results within a reasonable timeframe, even at the expense of sacrificing some performance.

\subsection{Satellite Matching Algorithm}

The \name algorithm, designed in this work, is a heuristic approach based on the greedy strategy specifically tailored to address this problem, which embodies the greedy approach in two key aspects: firstly, in selecting access satellites for each edge cloud by arranging them in descending order of data volume, and secondly, in defining bandwidth levels by selecting the nodes with the minimum potential connectivity among satellites with the highest available bandwidth levels. Arranging edge clouds in descending order of data volume ensures that bandwidth resources are allocated preferentially to nodes requiring higher data transmission, thereby reducing bandwidth contention among nodes. Similarly, prioritizing nodes with the minimum potential connectivity helps alleviate competition during the selection of access nodes for subsequent edge clouds.

\begin{figure*}[!t]
\centering
\subfloat[Transmission duration results]{\includegraphics[width=2.2in]{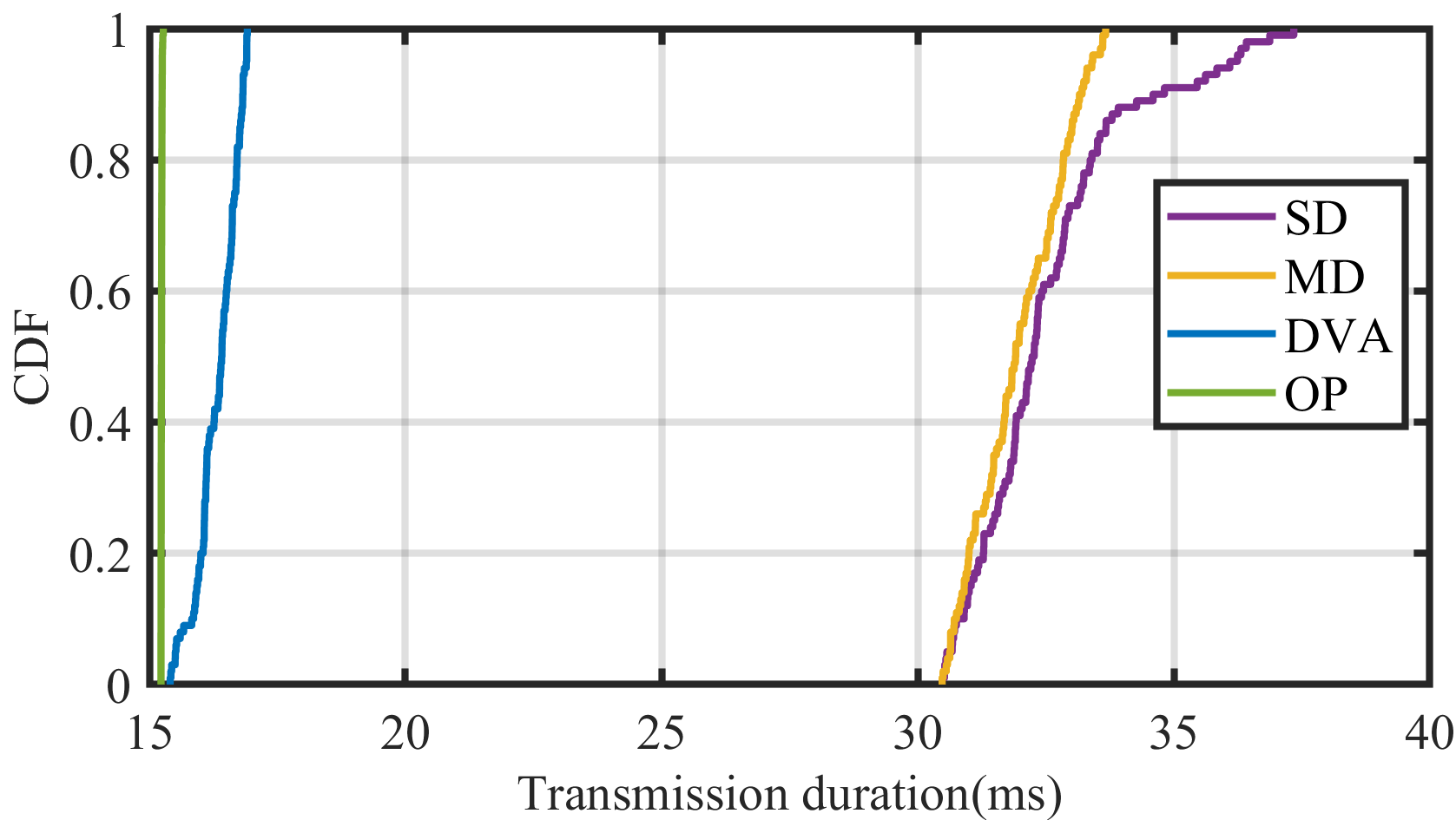}}
\hfil
\subfloat[Access network bandwidth results]{\includegraphics[width=2.2in]{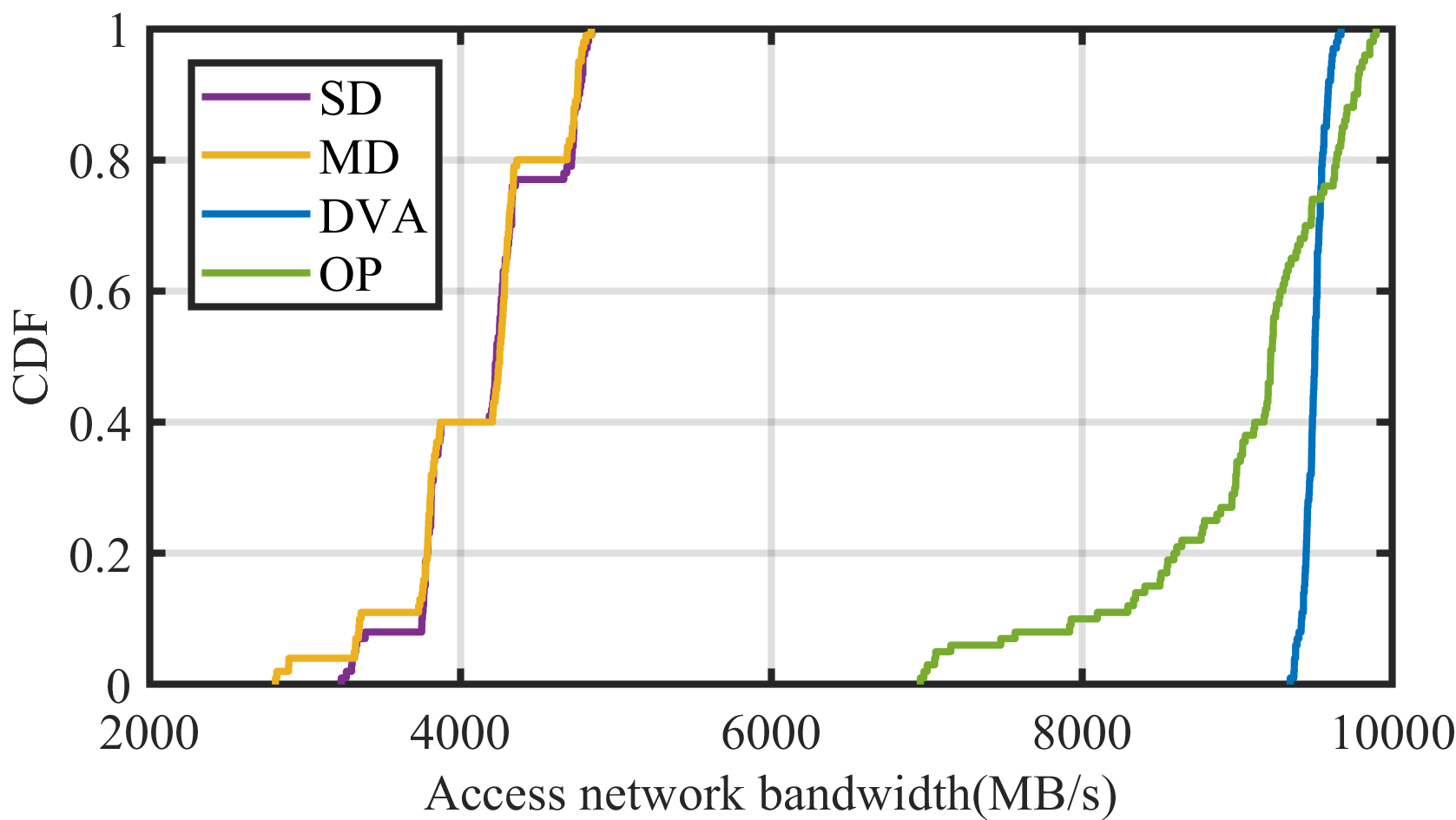}}
\hfil
\subfloat[Calculation finish time]{\includegraphics[width=2.2in]{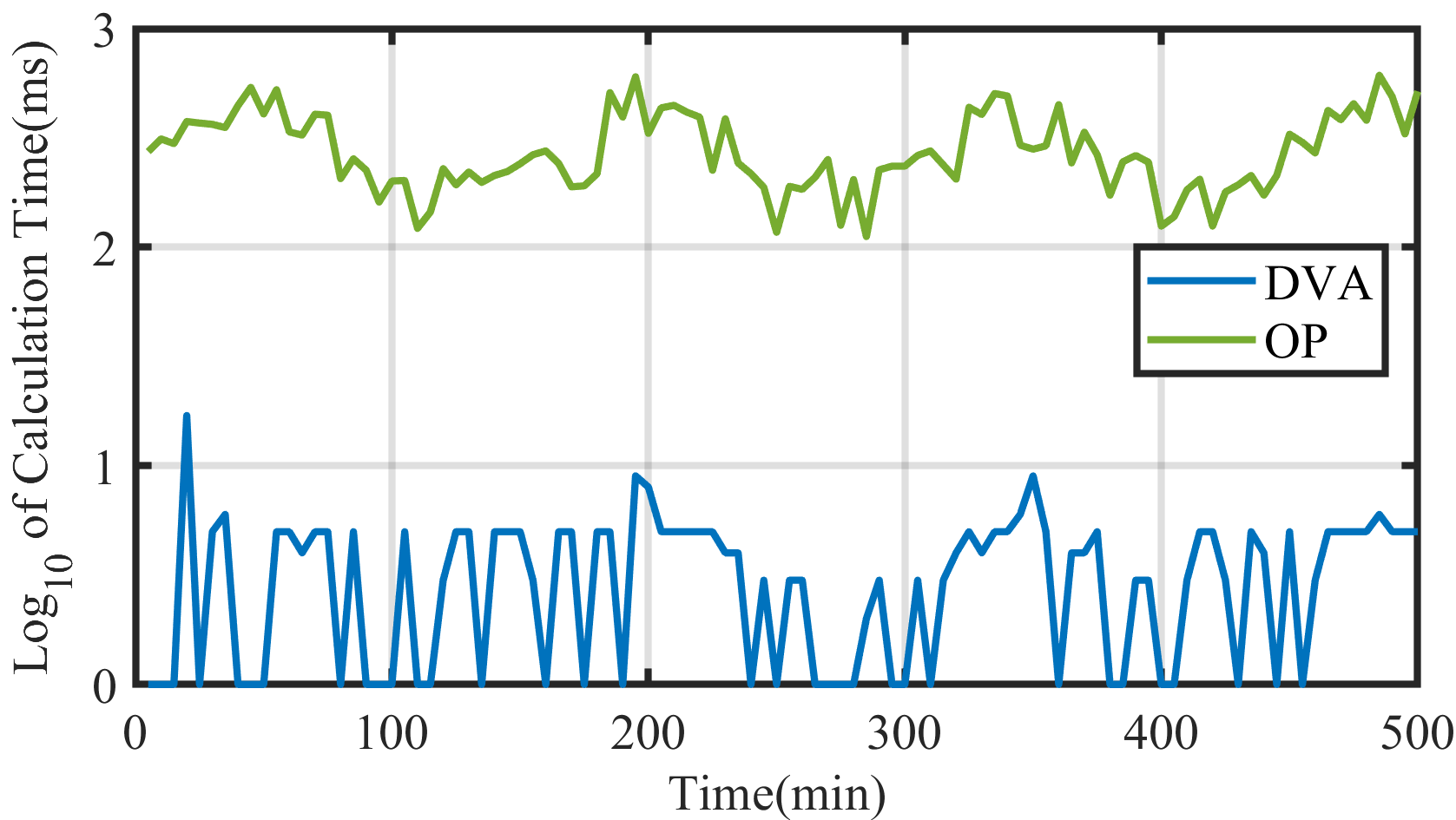}}
\caption{Performance Test Results of Four Algorithms}
\end{figure*}

The process of matching access satellites can be outlined as follows. Firstly, the edge cloud node set \( E \) needs to be sorted in descending order based on data volume. Then, the matching process begins by selecting satellites in descending order of data transmission size. For an edge node, bandwidth levels for candidate satellites are determined based on the data transmission size. For instance, if there is a need to transmit 50MB of data, the bandwidth level for candidate satellites is determined as 50MB/s. Following this, among a group of satellites with the highest available bandwidth level, the node with the fewest potential connections is selected as the access satellite.

After determining the access satellite, it is necessary to subtract the data volume of the edge cloud from the available bandwidth of that satellite. Additionally, the potential connections of other access satellites need to be updated by subtracting 1 from the potential connections of all candidate access satellites for that edge cloud. Following this principle, iterate through each edge cloud to obtain the final selection results for access satellites.

\begin{algorithm} 
\caption{\name Algorithm} 
    \begin{algorithmic}
    \renewcommand{\algorithmicrequire}{\textbf{Input:}}
    \renewcommand{\algorithmicensure}{\textbf{Output:}}
    \REQUIRE Edge list $lE$, Satellite list $lS$, Relationship map $mE2S$,~$mS2E$
    \ENSURE Match result map $mR$
    \STATE SortByDataVolumeDescending($lE$);
    \FOR { $e$ ~in~ $lE$ }
        \STATE CandidateSatelliteList $lCS$, AccessSatellite $AS$;
        \STATE $lCS \gets$ HighestBandwidthLevel($mE2S[e]$);
        \STATE $AS \gets$ MinimumPossibleConnectedEdges($lCS$);
        \STATE add $AS$ to $mR$;
        \STATE UpdateAvailableBandwidth($AS$);
        \FOR{ $s$ ~in~ $mE2S[e]$ }
            \STATE PossibleConnectedEdgesMinusOne($s$);
        \ENDFOR
    \ENDFOR
    \end{algorithmic} 
\end{algorithm}

Based on these two simple principles, it is possible to fully exploit the available bandwidth of the access network while avoiding extensive enumeration operations. This reduces the time complexity from \(O(m^n)\) to \(O(m \times n)\), enabling real-time initiation of the data transmission process. Due to the predictability of satellite trajectories and transmission times, the same greedy principle can be applied for necessary access satellite switching. This involves reselecting satellites for the nodes or node sets requiring a switch. Here is the pseudo code for the algorithm part.

\section{Evalutaion} \label{sec:evaluation}
In this section, we will evaluate the performance of \name in a simulated environment. This evaluation will include testing the duration of edge cloud data transmission, the throughput of the access network, and the computation time for calculating matching results. Additionally, we conducted tests on different configurations of satellite constellations to validate the robustness of the algorithm.

% \subsection{Testing Data and Environment}
\subsection{\newrev{Experimental Setup}}
The edge cloud data is sourced from the global edge nodes of Amazon CloudFront~\cite{kataru2023cost}. CloudFront is a global network comprising over 600 access points, covering more than 100 cities in over 50 countries and regions worldwide. For the data volume on each edge node, we will estimate and simulate based on the local population of the site, assuming a user base of 1\% of the population and with an average of 0.1 KB of data transmission per user.

We employ the Starlink Shell I constellation as our satellite constellation~\cite{yin2021beam}, with each satellite's total uplink bandwidth set to 500 MB/s. We utilize the mainstream satellite simulation tool, Systems Tool Kit (STK), to simulate satellite motion trajectories. We selected the Shortest Path~(SP) algorithm based on the shortest distance and the Maximum Duration~(MD) algorithm based on the longest visible time as benchmarks for access satellite selection~\cite{liu2022enabling}. Additionally, we chose an Integer Linear Programming (ILP) algorithm solved using the Gurobi optimizer as the optimal solution (OP). 

Under the same network conditions, we selected 20 CloudFront nodes located in the North American region for a simulation process lasting 24 hours. Eventually, we sampled the constellation's status at intervals of 5 minutes, totaling 100 instances, along with the same random background traffic, to serve as inputs for algorithm comparison, evaluating their performance.

\subsection{\newrev{Overall Performance}}

The performance metrics measured for the algorithms generally include the following aspects.

\textbf{Transmission Duration Analysis.} We first compared the transmission duration of the access network. As shown in Fig.~4(a), under the same conditions, \name achieved an average reduction of 49.7\% and 48.8\% in transmission time compared to the SP and MD algorithms, respectively. This demonstrates a significant improvement in performance. Thanks to the greedy approach in the design, \name effectively mitigates the occurrence of multiple edge clouds simultaneously selecting the same access satellite within their line of sight. Meanwhile, compared to the theoretical optimal result, \name only incurred an increase of approximately 8\% in transmission duration. Therefore, \name effectively matches more suitable access satellites based on data volume and satellite available bandwidth, thereby reducing the transmission duration of the access network.

\textbf{Access Network Throughput Analysis.} Fig.~4(b) illustrates the comparison of the achievable access network throughput for each algorithm. Compared to SP and MD, the \name algorithm improved the average access network throughput by 2.28 times and 2.30 times, respectively. Due to the consideration of transmission duration rather than larger available bandwidth in the OP model, \name's available access network throughput is slightly higher than that of OP, reaching 1.07 times that of OP. This indicates that the matching strategy based on greedy principles in \name avoids unnecessary bandwidth contention, particularly in cases where multiple edge clouds can observe the same access satellite, resulting in a more significant effect.

\textbf{Computation Duration Analysis.} Finally, we compared the computation duration of the \name algorithm with the runtime of the linear programming algorithm using the Gurobi optimizer. As shown in Fig.~4(c), due to the significant reduction in the search scope of the solution space enabled by \name, under the same computational conditions, the computation duration of the OP algorithm is approximately 290 ms, while the runtime of the \name algorithm consistently remains below 1 ms, significantly lower than the time required by the linear programming algorithm using enumeration methods. Such computation duration is similar to that of the SP and MD algorithms in practical applications, meeting the real-time requirements of access satellite scheduling.

\subsection{Performance Among Different Constellations}

\begin{table}[!t]
\caption{Satellite Constellation Parameters}
\label{tab:table1}
\centering
\begin{tabular}{|c|c|c|c|c|c|c|c|}
\hline
Name & Telesat-Inclined & OneWeb & Starlink Shell-1 \\
\hline
Orbits & 5 & 18 & 66 \\
\hline
Satellite per orbit & 10 & 40 & 24 \\
\hline
Altitude & 1200 & 1200 & 550 \\
\hline
Inclination & 34.7 & 87.9 & 53 \\
\hline
Phase shift & 0 & 0 & 1 \\
\hline
Min elevation & 20 & 55 & 25 \\
% \hline
% Half cone & 52.2558 & 28.8596 & 56.5 \\
\hline
\end{tabular}
\end{table}

\begin{figure}[!t]
\centering
\includegraphics[width=2.3 in]{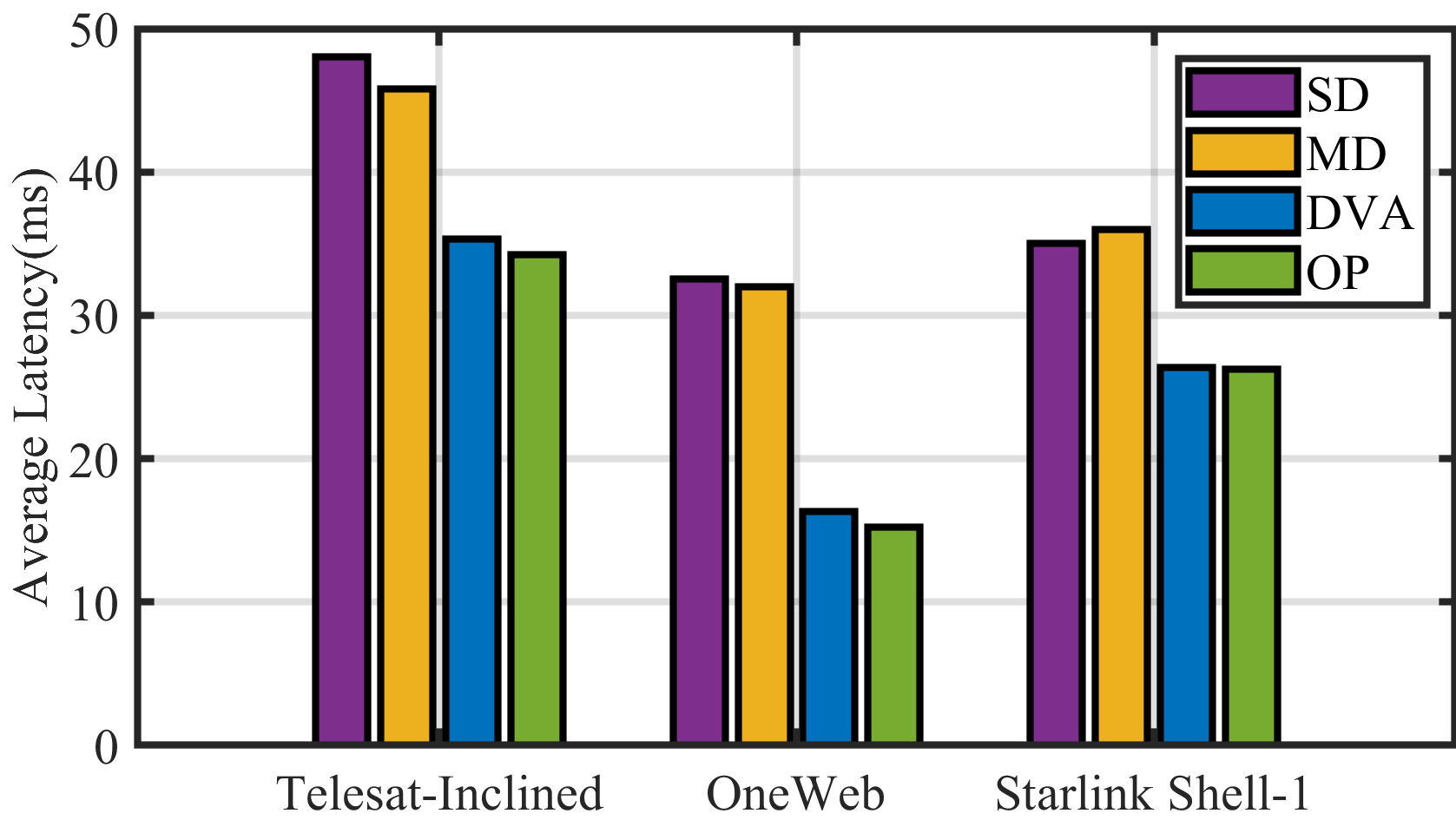}
\caption{The performance results among different constellations.}
\end{figure}

After completing the basic performance tests, we tested the running effects of algorithms in different configurations of constellations under the same edge cloud deployment conditions~\cite{10494942}. The relevant parameters of the three constellations are shown in Table~\ref{tab:table1}. As shown in Fig. 5, for the Telesat-Inclined, OneWeb, and Starlink Shell-1 constellations, the average access network duration of \name was significantly lower than the two default constellation selection algorithms, and approached the theoretical optimum value. This further demonstrates the effectiveness of the algorithm in this scenario, as it can adapt to different constellations, consistently yielding superior constellation selection results.

\section{Conclusion} \label{sec:conclusion}
In order to address the issue of transmitting distributed data from edge clouds through satellite networks to centralized data centers for processing, we have devised the \name algorithm. Based on greedy principles, this algorithm matches access satellites for each edge cloud, thereby avoiding unnecessary bandwidth contention, reducing transmission completion latency, and increasing access network throughput. Compared to the default access satellite selection algorithm, the transmission time has been reduced by approximately 50\%, while the access network throughput has increased by over 2 times. Compared to the theoretical optimal algorithm, under the condition of increasing the access network transmission duration by less than 8\%, the computation time has been significantly reduced from 290 ms to within 1 ms. As a potential future direction, we are looking forward to extending our method to improve the performance of various applications such as distributed learning systems~\cite{lin2024efficient,lin2024adaptsfl,lin2024split}, ISAC~\cite{hu2023holofed}, large language models~\cite{fang2024automated,lin2023pushing}, etc in LEO satellite networks.

\bibliographystyle{IEEEtran}
\bibliography{ref}

\end{document}